# Registration-Guided Deep Learning Image Segmentation for Cone Beam CT–based Online Adaptive Radiotherapy

Lin Ma[1,*], Weicheng Chi[1,2,*], Howard E. Morgan[1], Mu-Han Lin[1], Mingli Chen[1], David Sher[1], Dominic Moon[1], Dat T. Vo[1], Vladimir Avkshtol[1], Weiguo Lu[1†] and Xuejun Gu[1†]

[1]Medical Artificial Intelligence and Automation Laboratory, Department of Radiation Oncology, University of Texas Southwestern Medical Center, 2280 Inwood Rd, Dallas, TX 75390, USA

[2]School of Software Engineering, South China University of Technology, Guangzhou, Guangdong, 510006 China

[*]These authors contributed equally to this work.

[†]Authors to whom any correspondence should be addressed.

[†]Corresponding Author: Xuejun.Gu@UTSouthwestern.edu Weiguo.Lu@UTSouthwestern.edu

## Abstract

Adaptive radiotherapy (ART), especially online ART, effectively accounts for positioning errors and anatomical changes during the course of treatment. One key component of online ART processes is accurately and efficiently delineating organs at risk (OARs) and targets on online images, such as Cone Beam Computed Tomography (CBCT), to meet the online demands of plan evaluation and adaptation. Deep learning (DL)-based automatic segmentation has gained great success in segmenting planning CT, but its applications to CBCT yielded inferior results due to the low image quality and limited available contour labels for training. To overcome these obstacles to online CBCT segmentation, we propose a registration-guided DL (RgDL) segmentation framework that integrates image registration algorithms and DL segmentation models. The registration algorithm generates initial contours, which were used as guidance by DL model to obtain the accurate final segmentations. We had two implementations the proposed framework—Rig-RgDL (Rig for rigid body) and Def-RgDL (Def for deformable)—with rigid body (RB) registration or deformable image registration (DIR) as the registration algorithm respectively and U-Net as DL model architecture. The two implementations of RgDL framework were trained and evaluated on seven OARs in an institutional clinical Head and Neck (HN) dataset. Compared to the baseline approaches using the registration or the DL alone, RgDL achieved more accurate segmentation, as measured by higher mean Dice similarity coefficients (DSC) and other distance-based metrics. Rig-RgDL achieved a DSC of 84.5% on seven OARs on average, higher than RB or DL alone by 4.5% and 4.7%. The DSC of Def-RgDL is 86.5%, higher than DIR or DL alone by 2.4% and 6.7%. The inference time took by the DL model to generate final segmentations of seven OARs is less than one second in RgDL. The resulting segmentation accuracy and efficiency show the promise of applying RgDL framework for online ART.

## 1 Introduction

Radiotherapy (RT) has long played an important role in cancer care and has steadily advanced with new technological developments. Modern RT is often conducted with intensity modulation (IMRT) techniques. IMRT treatment plans have traditionally been based on computed tomography (CT) simulation scans done approximately 1-2 weeks prior to the start of therapy, which inherently assumes that anatomy will not change during the treatment course. However, it is well known that both gross tumor volume (GTV) and organs at risk (OARs) change size and shape during the course of treatment. For example, during head and neck (HN) radiotherapy, parotid glands and submandibular glands (SMGs) decrease in size even as early as week 2 of treatment, and many patients also experience weight loss resulting in reduced neck circumference[1, 2]. Taken together, these anatomic changes may lead to unintended overdosage to OARs and/or underdosage to targets[3]. Recently, adaptive radiotherapy (ART), an innovative RT modality, has garnered interest for its potential to improve radiotherapy outcomes by adapting plans to up-to-date anatomy. Several centers have developed offline approaches to identify and adapt patients' treatment plans to their evolving anatomy[4-8]; however, none have gained widespread adoption yet because of the time-consuming process of re-contouring and re-planning required to make this practical. In contrast to offline ART, online ART involves online plan adaptation using images acquired at treatment position, most commonly with online Cone Beam CT (CBCT). Given that online ART would occur while patients lie on the treatment couch, online ART would be even more time-sensitive and resource-demanding. Therefore, to make online ART feasible, it is crucial to develop a fast and accurate CBCT-based automatic segmentation tool to streamline this process.

Deep learning (DL) is a promising technique for fast and accurate automated image segmentation. U-Net[9] was firstly constructed for biomedical image segmentation. In view of U-Net's ability to link low-level and high-level contexts, researchers have widely adopted its canonical and derivative networks for RT image segmentation and have achieved promising results with various RT image modalities, such as magnetic resonance imaging (MRI) and CT[10-13]. However, automatic contouring on CBCT via a direct DL approach confronts several challenges: 1) inferior image quality—compared to MRI and CT, CBCT has more noise, artifacts and lower soft tissue contrast; and 2) limited available contour labels—as CBCT is mainly designated for image-guided treatment delivery, clinicians rarely delineate structures on CBCT, which precludes the easy acquisition of large amounts of data. Thus, it is anticipated that DL-based segmentation networks trained directly on CBCT images with limited labels will yield disappointing performance for automatic contouring.

Integrating prior knowledge into DL segmentation model has been demonstrated to effectively address the aforementioned challenges. One solution is the self-supervised learning strategy, which used pre-trained tasks to generate prior knowledge for organ segmentation[14, 15]. However, they cannot explain the relationship between the pre-trained tasks and the task of segmentation explicitly. In contrast, positional information of organs is prior knowledge to segmentation model. Direct informing the model of positional information is a more straightforward and explainable way. Chen[13] proposed a recursive ensemble organ segmentation (REOS) framework, which first localized regions of interest (ROI) to crop a local patch of each head and neck (HN) organ for final auto-contouring. Kaul[16] designed the FocusNet, which delineated OARs in a global segmentation model and fine-tuned the rough segmentation results of small organs in designated branches. These methods require auxiliary networks for localization to segment the organs of interest, thus increasing model parameters (complexity) and consequently decreasing auto-segmentation efficiency. For online OAR segmentation on CBCT, planning CTs and their corresponding contours can provide the lacking positional information. Specifically, image registration methods, such as deformable image registration (DIR), which constructs the voxel-to-voxel correspondence between planning CT and daily images, is the working horse of online segmentation. They can propagate planning contours to daily megavoltage CT (MVCT)/CBCT images[17-19], though the accuracy of the propagated contours is limited by the inferior image quality of online MVCT/CBCT.

In this study, we attempt to achieve accurate and rapid HN OAR segmentation on CBCT for online ART by integrating image registration with DL segmentation. We propose a registration-guided DL (RgDL) segmentation framework: the transformed/deformed masks (propagated contours) initially generated by registration algorithm—providing positional guidance—are input to a DL model along with CBCT for final accurate segmentation. RgDL framework can improve the accuracy of DL segmentation by incorporating the prior knowledge of planning contours brought through registration. Viewed in another way, it can be regarded as refining registration-propagated contours to the online CBCT image via a DL model. In current CBCT-based online ART workflows, OARs are delineated by registration-based contour propagation, then contours are edited manually. RgDL can fit into clinical ART workflow by adding a DL model after registration, which aims to refine the propagated contours and lessen the workload of manual contour editing.

The rest of the paper is organized as follows. Section 2 details the implementation of RgDL and the experiments carried out to evaluate its segmentation accuracy. Sections 3.1 reports the quantitative segmentation results for the RgDL implementations and compares them with those of baseline methods: registration and DL. Section 3.2 investigates the influence of training dataset size on the accuracy of the RgDL framework. Section 4 discusses the clinical applicability and directions for future research.

## 2    Materials & Methods

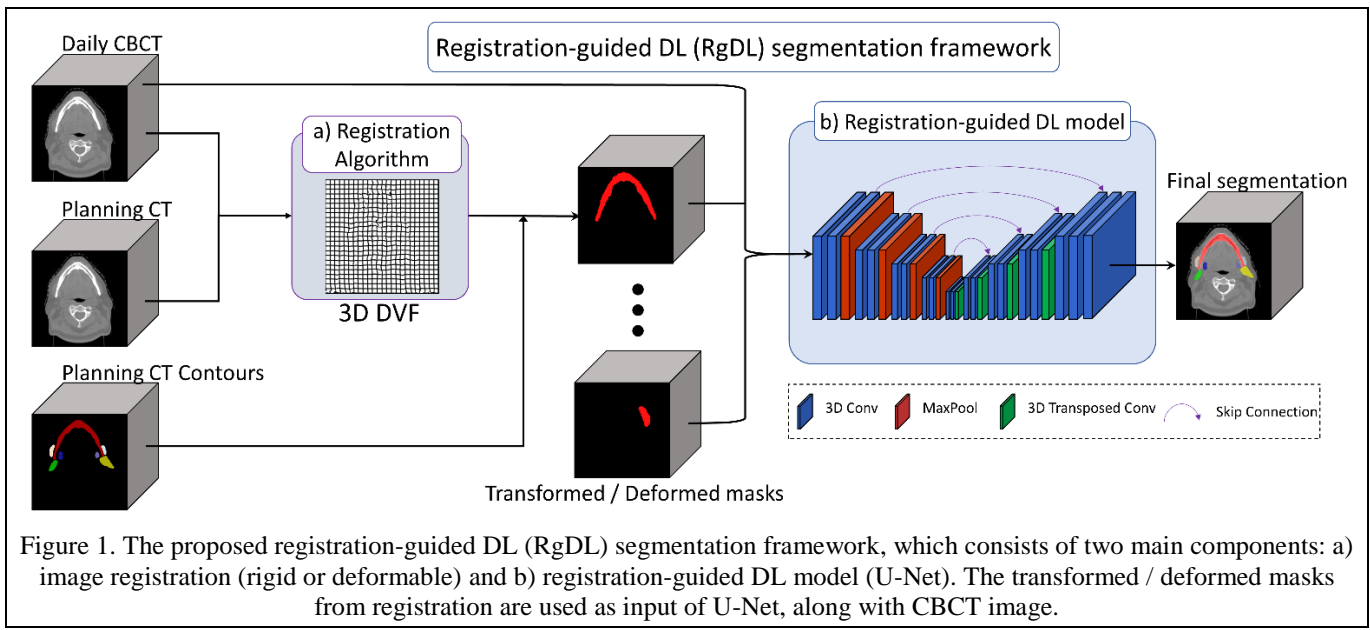

Figure 1. The proposed registration-guided DL (RgDL) segmentation framework, which consists of two main components: a) image registration (rigid or deformable) and b) registration-guided DL model (U-Net). The transformed / deformed masks from registration are used as input of U-Net, along with CBCT image.

### 2.1    Registration-guided DL (RgDL) segmentation framework

The proposed RgDL segmentation model consists of two main steps: registration-based propagation of planning contours and DL-based segmentation (figure 1). In the former, the propagated contours (transformed/deformed masks) are generated from the planning contours by registering the planning CT to the daily CBCT; in the latter, the propagated contours serve as model input and guide the DL model to segment accurately. We have two implementations of RgDL with different registration algorithms, and the DL models were trained separately for each type of registration guidance.

### 2.1.1    Image registration algorithms

We have employed two different registration algorithms for RgDL, rigid body (RB) registration and deformable image registration (DIR). The goal of registration is to find a voxel-to-voxel correspondence from moving image $I_m$ to fixed (target) image $I_f$. The implementation details of two algorithms are described in this subsection. The results of the two RgDL implementations will be reported and compared in the Results section.

The RB registration transforms the moving image to the fixed image by rigid body transform, which has six degrees of freedom (three for translation and three for rotation). We adopted ITK4.13 (Kitware and Insight software consortium) multi-resolution registration method with Mattes mutual information similarity measure and regular-step gradient-descent optimizer for RB registration[20]. Optimization process is stopped when the change is less than 0.001cm or 0.01 degree.

DIR deforms the image by a parameter-controlled warping model or by a free-form deformation defined by a voxelized deform vector field (DVF) $\Phi$. We adopted Demons[21], a free-form deformation algorithm, as the DIR for the second implementation of RgDL. It aims to find an optimal DVF $\Phi$ by maximizing the similarity between the two images:

$$\widehat{\Phi} = arg\max_{\Phi} S(\mathcal{T}(I_m, \Phi),\ I_f) \tag{1}$$

where the similarity metric $S(\cdot)$ is cross correlation and $\mathcal{T}(I_m, \Phi)$ is the transformation of the moving image with the DVF $\Phi$. DVF is optimized by DIR force, which is mainly driven by the intensity gradient of the fixed image, as follows:

$$\Phi^{(n)} = \frac{\left(I_m^{(n-1)} - I_f\right)\nabla I_f}{\left|\nabla I_f\right|^2 + \left(I_m^{(n-1)} - I_f\right)^2} \tag{2}$$

where $\Phi^{(n)}$ and $I_m^{(n-1)}$ are the DVF of the $n^{th}$ iteration and the warped moving image of the previous iteration, respectively. Moreover, a Gaussian filter is applied to the DVF in each iteration for regularization and smoothness. The stopping criterion is $l^{(n-10)} - l^{(n)} \leq 1.0 \times 10^{-4}$, where $l^{(n)} = \frac{\sum\left|d\Phi^{(n+1)}\right|}{\sum\Phi^{(n)}}$, because this measure has a closer correspondence with spatial accuracy than correlation coefficient[21]. We run Demons DIR after the moving image is rigidly aligned to the fixed image. The Demons algorithm is implemented by our in-house GPU-accelerated Demons package with simultaneous intensity correction tailored for CT to CBCT deformable registration[21, 22].

In this study, planning CT is the moving image and daily CBCT is the fixed image. Planning CT contours are propagated to daily CBCT by the translation and rotation operations obtained by RB registration or the DVF optimized by DIR. The resulting propagated contours are input into DL model as binary masks for each OAR.

### 2.1.2    Registration-guided DL model

We employed U-Net as the backbone of DL model architecture. Our model comprises a contracting path and an expanding path to extract both the high-level and the low-level context for the final segmentation. The contracting path acts as an encoder with four convolutional blocks, each followed by a 3D max-pooling layer with a kernel size of 2. The output channel of the first convolutional layer in the contracting path is 32, and we double the number of the convolutional layers that follow every max-pooling layer. Before the first convolutional block of each level in the expanding path, skip connections are applied to concatenate the feature maps from the contracting path of the same level to combine high-level and low-level feature representations. In the expanding path, we exploit four convolutional blocks symmetric to the contracting path to up-sample feature maps and halve the number of channels of convolutional layers after each transposed convolutional layer. After the expanding path, we add an extra convolutional layer and a softmax function to convert the feature maps to the organ probability maps. All of the convolutional blocks consist of two consecutive convolutional layers of $3 \times 3 \times 3$ kernels, each followed by a batch normalization layer and a rectified linear unit (ReLU) activation function.

To integrate registration-generated contour information into DL model, we utilize the CBCT and transformed / deformed masks (propagated contours) as multi-channel inputs to the DL segmentation model (figure 1.b). To preserve the physics meaning of propagated contours, we create binary mask maps for each OAR individually. For example, with seven OARs in our study case, we created seven maps of binary intensity with 0 outside the OAR contours and 1 inside. Supervised by the ground truth (GT) contours that experts segment on CBCT, DL model learns to adapt the propagated contours to CBCT anatomy. The model outputs eight channels of probability masks, comprising background (non-OAR) and seven OARs. The predicted binary OAR masks are obtained by performing a voxel-wise classification. The class (background or one of the seven OARs) with the highest probability is assigned to each voxel. The DL model is optimized by minimizing the volumetric soft Dice loss function[10], which improves the ratio of intersection over union.

## 2.2    Experimental data

To evaluate the proposed method, we created an in-house HN dataset. It contains the imaging and planning information of 37 patients with HN cancer treated in our institution. Participants included had histologically confirmed locally advanced HN cancer, had received definitive chemoradiotherapy (69.96-70 Gy in 33-35 fractions), and had daily CBCT images acquired for treatment setup. Most primaries tumors were oropharyngeal (n=35), with the remaining being supraglottic (n=2). Each patient case collected contains a planning CT, planning contours, and an online CBCT that was acquired at least four weeks after the planning CT. The corresponding CBCT contours were manually delineated in Velocity™ by one oncologist and used as GT for this research. CT scans were acquired on a Philips 16-slice Brilliance large-bore CT simulator. The in-plane resolution of the CTs varies between 1.17 mm and 1.37 mm, and the slice thickness is 3 mm. CBCT scans were acquired by the onboard imager of Varian TrueBeam™ (Varian Medical Systems, Palo Alto, CA) machines with 100 kVp, 150 mAs and full-fan 200° acquisition. The voxel resolution of the CBCT is 0.51 mm × 0.51 mm × 1.99 mm. The imaging field of CBCT consistently covers skull base to thoracic vertebrae in the superior-inferior direction.

This study focuses on seven HN OARs: mandible, left & right parotid glands, left & right submandibular glands (SMG), and left & right masseters. The 37 patients were randomly split into training (22), validation (7) and test (8) datasets. Any given patient may not have the complete set of seven OARs due to surgical resection or tumor encroachment. The numbers for each organ in each subset are shown in table 1. For a fair comparison, the test data were only accessed during the final evaluation of segmentation performance, and the validation set was used to tune the network's hyper-parameters and to select the best model.

Table 1. The numbers of patients (CBCT images) and organs in the training, validation, and test sets.

| Dataset | #pts | Mandible | Parotid-R | Parotid-L | SMG-R | SMG-L | Masseter-R | Masseter-L |
|---------|------|----------|-----------|-----------|-------|-------|------------|------------|
| Training | 22 | 22 | 19 | 19 | 19 | 17 | 14 | 13 |
| Validation | 7 | 7 | 7 | 5 | 7 | 7 | 4 | 4 |
| Test | 8 | 8 | 7 | 7 | 7 | 8 | 7 | 7 |

## 2.3 Experimental design

We have two implementations of RgDL segmentation framework—Rig-RgDL (*Rig for Rigid body*) and Def-RgDL (*Def for Deformable*). The Rig-RgDL employed RB registration to propagate contours, and the DL model was trained only by RB-propagated contour input. The Def-RgDL used Demons DIR, and the DL model was trained only by DIR-generated contour input. We evaluated the Rig-RgDL and Def-RgDL on our in-house HN dataset. They are compared to three baseline algorithms: RB registration, DIR (Demons) and the DL model itself (without registration guidance), whose input contains only one channel for CBCT. The three baselines are chosen because they are individual components of RgDL. We used the Dice similarity coefficient (DSC—two folds of intersection divided by union), 95th percentile Hausdorff distance $(HD_{95})$[11] and average symmetric surface distance $(ASD)$[11] to evaluate the segmentation performances of two RgDLs (Rig-RgDL and Def-RgDL) and three baseline algorithms (RB, DIR and DL).

Experiments were conducted on a Windows 10 (x64) workstation with an NVIDIA GeForce 2080 Ti GPU with 12 GB memory, and DL model was implemented with Python (version 3.7.3) and library Pytorch (version 1.6.0). The model input, registration-propagated contours (transformed/deformed masks), were generated in the following steps. Firstly, we registered the planning CT volume to the CBCT volume (512 $\times$ 512 $\times$ 93 with resolution 0.51 mm $\times$ 0.51 mm $\times$ 1.99 mm) by RB or DIR and propagated planning contours. Secondly, the superior most 64 CBCT slices were extracted—as all seven OARs only reside in this range—and then down-sampled by a factor of four in axial plane. The resulting 128 $\times$ 128 $\times$ 64 volumes served as inputs for the U-Net, corresponding to a size of 26.1 cm $\times$ 26.1 cm $\times$ 12.7 cm with a voxel resolution of 2.04 mm $\times$ 2.04 mm $\times$ 1.99 mm. The DL models in the two implementations of RgDL (Rig-RgDL and Def-RgDL) were trained separately, with the same GT but different registration guidance inputs (transformed or deformed masks). We trained the DL models with the Adam optimizer and an initial learning rate of $10^{-3}$ for 100 epochs. The learning rate was reduced by 10 after 50 and 80 epochs to arrive at the local minimum. The best model on the validation set was used for the final evaluation on the test sets.

## 3 Results

### 3.1 Segmentation accuracy of RgDL framework

The segmentation accuracy metrics (DSC, $HD_{95}$ and ASD) of two implementations of RgDL (Rig-RgDL and Def-RgDL) and three baseline methods (RB, DIR and DL) are reported in table 2. Rig-RgDL achieved a mean DSC of 84.5% for seven OARs on average, outperforming RB contour propagation and DL segmentation without guidance by 4.5% and 4.7%, respectively. The DSC of Def-RgDL is better than DIR and DL by 2.4% and 6.7%. We can also observe the tendency that a more accurate registration algorithm lead to a more accurate implementation of RgDL. The DIR is better than RB by 4.1% DSC, 0.5mm $HD_{95}$ and 0.3mm ASD. Accordingly, Def-RgDL is better than Rig-RgDL by 2% DSC, 0.4mm $HD_{95}$ and 0.14mm ASD. For all five methods, the segmentation accuracy on mandible and masseter are better than parotids and SMGs, as bony and muscular structures have better contrast than glandular structures on CBCT.

Table 2. Segmentation accuracy of proposed methods (Rig-RgDL and Def-RgDL) and baseline methods (DL, RB and DIR). One metric is reported in one block. Higher DSC and smaller HD$_{95}$ and ASD values indicate better segmentation accuracy. Results of five methods are evaluated on the test dataset (table 1). The best results for each OAR are highlighted in bold font.

| Metric | Methods | Mandible | Parotid-R | Parotid-L | SMG-R | SMG-L | Masseter-R | Masseter-L | Mean |
|---|---|---|---|---|---|---|---|---|---|
| DSC (%) | DL | 88.7±1.8 | 80.6±4.5 | 78.7±5.0 | 71.1±6.5 | 70.1±6.1 | 86.0±3.7 | 83.5±8.0 | 79.8 |
| | RB | 83.8±5.8 | 79.5±7.0 | 80.3±3.7 | 72.2±9.9 | 70.8±8.6 | 85.8±3.5 | 87.5±3.3 | 80.0 |
| | DIR | 90.7±1.8 | 86.4±2.4 | 85.6±3.4 | 73.4±11.5 | 76.2±3.4 | 86.8±2.9 | 89.5±1.6 | 84.1 |
| | Rig-RgDL | 88.2±2.1 | 86.4±3.1 | 85.2±3.6 | 76.1±8.3 | 75.9±4.2 | 88.6±2.9 | 90.8±2.3 | 84.5 |
| | Def-RgDL | **92.8±1.3** | **88.6±1.3** | **86.4±3.8** | **76.6±10.1** | **81.4±1.5** | **88.6±2.6** | **91.4±2.2** | **86.5** |
| HD$_{95}$ (mm) | DL | 3.36±0.81 | 9.20±4.05 | 7.49±1.98 | 5.67±2.24 | 6.47±2.97 | 3.25±1.06 | 4.20±1.62 | 5.66 |
| | RB | 2.58±0.74 | 4.01±1.65 | 3.94±1.02 | 3.74±1.15 | 4.28±1.09 | 2.34±0.72 | 2.46±0.73 | 3.34 |
| | DIR | 2.00±0.02 | 3.01±0.92 | 2.84±0.85 | 3.18±1.07 | 3.46±0.56 | 2.75±1.11 | 2.29±0.38 | 2.79 |
| | Rig-RgDL | 4.13±1.86 | 2.75±0.89 | 2.74±0.82 | **2.97±0.81** | 3.72±1.03 | **2.16±0.28** | 2.16±0.28 | 2.95 |
| | Def-RgDL | **1.99±0.01** | **2.64±0.86** | **2.58±0.72** | 3.05±1.17 | **2.87±0.61** | 2.60±0.87 | **2.04±0.01** | **2.54** |
| ASD (mm) | DL | 0.51±0.15 | 1.56±0.52 | 1.55±0.40 | 1.66±0.75 | 1.68±0.34 | 0.72±0.26 | 1.05±0.50 | 1.25 |
| | RB | 0.64±0.32 | 1.30±0.58 | 1.18±0.36 | 1.26±0.50 | 1.42±0.44 | 0.73±0.18 | 0.61±0.19 | 1.02 |
| | DIR | 0.26±0.09 | 0.73±0.21 | 0.74±0.27 | 1.13±0.61 | 1.05±0.19 | 0.66±0.26 | 0.48±0.12 | 0.72 |
| | Rig-RgDL | 0.56±0.21 | 0.75±0.29 | 0.77±0.21 | 1.00±0.41 | 1.15±0.22 | **0.52±0.16** | 0.40±0.11 | 0.73 |
| | Def-RgDL | **0.20±0.04** | **0.59±0.14** | **0.69±0.22** | **0.96±0.54** | **0.79±0.09** | 0.56±0.20 | **0.35±0.08** | **0.59** |

The DSC of RgDLs and baselines on eight mandibles in test set (table 1) are plotted in figure 2.a&b. The DSC of two RgDL implementations on eight mandible cases are consistently better than baseline methods as 15 out of 16 data points are above the diagonal line. The segmentation results of RgDLs and baselines on one mandible case (arrowed in figure 2.a&b) are compared in figure 2.c-e. For this case, the accuracy of RgDLs in DSC (Rig-RgDL:85.3%, Def-RgDL:89.4%) are better than baselines (RB:76.9%, DIR:87.4%, DL:79.8%) with large margins, featuring by the positions of its data points (arrowed in figure 2.a&b) which are much higher than diagonal lines. It quantitatively demonstrates that RgDL can improve registration and DL. Due to the severe dental artifact, DL falsely segment teeth as mandible (figure 2.e). The guidance provided by registrations avoids this artifact-induced false segmentation (figure 2.c&d). Moreover, DIR (figure 2.c) provided more accurate guidance (green contours) than RB (figure 2.c) and Def-RgDL outperformed Rig-RgDL (yellow contours).

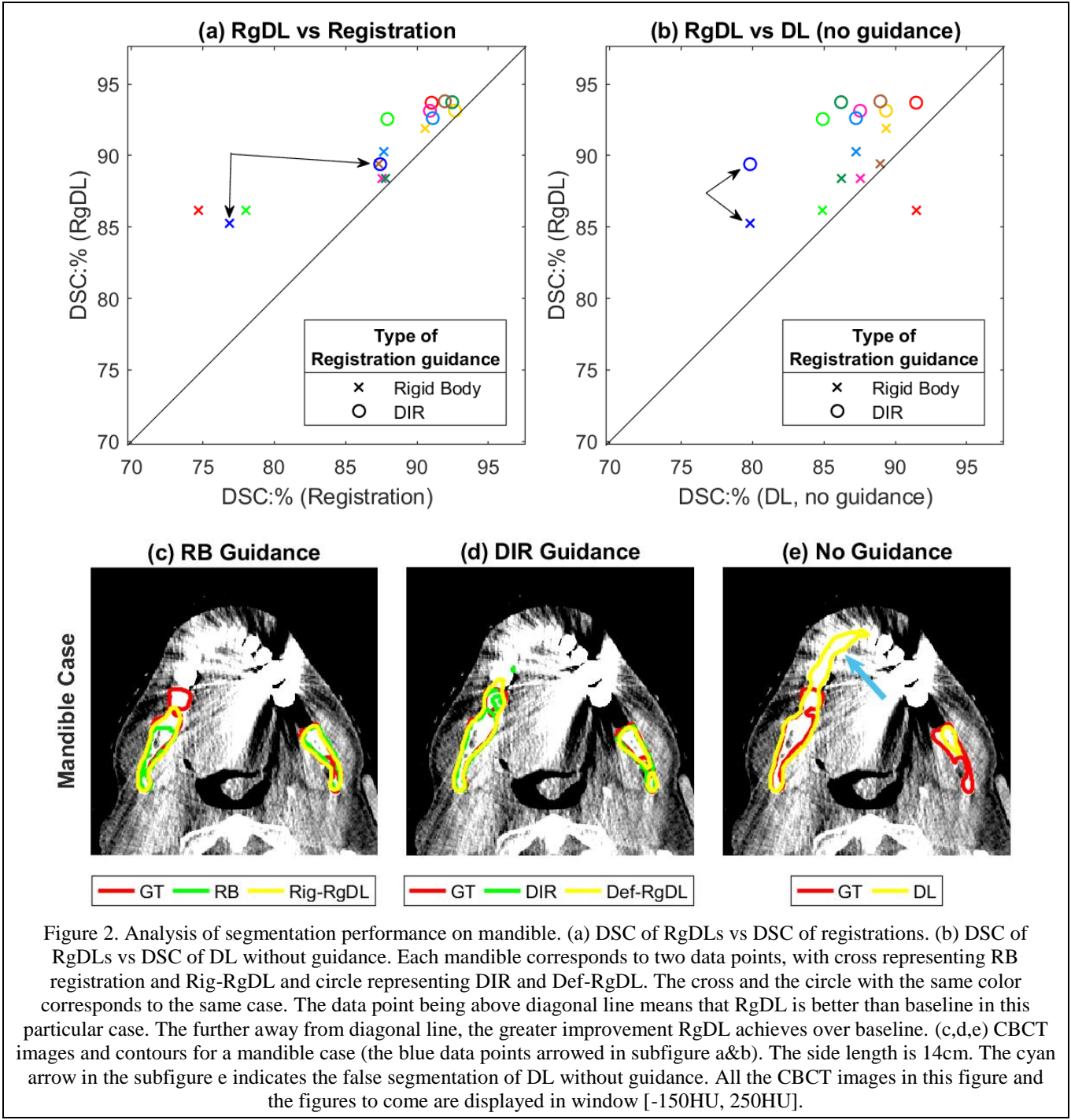

Figure 2. Analysis of segmentation performance on mandible. (a) DSC of RgDLs vs DSC of registrations. (b) DSC of RgDLs vs DSC of DL without guidance. Each mandible corresponds to two data points, with cross representing RB registration and Rig-RgDL and circle representing DIR and Def-RgDL. The cross and the circle with the same color corresponds to the same case. The data point being above diagonal line means that RgDL is better than baseline in this particular case. The further away from diagonal line, the greater improvement RgDL achieves over baseline. (c,d,e) CBCT images and contours for a mandible case (the blue data points arrowed in subfigure a&b). The side length is 14cm. The cyan arrow in the subfigure e indicates the false segmentation of DL without guidance. All the CBCT images in this figure and the figures to come are displayed in window [-150HU, 250HU].

The DSC of RgDLs and baselines on 14 parotids (left and right) in test set (table 1) are plotted in figure 3.a&b. The DSC of two RgDL implementations on 14 parotids are consistently better than baseline methods as 24 out of 28 data points are above the diagonal line. The segmentation results of RgDLs and baselines on one parotid case (arrowed in figure 3.a&b) are compared in figure 3.c-e. For this case, the accuracy of RgDLs in DSC (Rig-RgDL:86.8%, Def-RgDL:91.8%) are better than baselines (RB:75.9%, DIR:85.8%, DL:79.3%) with large margins, quantitatively demonstrating that RgDL can improve registration propagation and DL segmentation. The orange arrow in figure 3.c illustrates the tissue shrinkage or retraction due to treatment. Rig-RgDL can rectify the RB propagated contour by the CBCT image in the places pointed by orange arrow and cyan arrow.

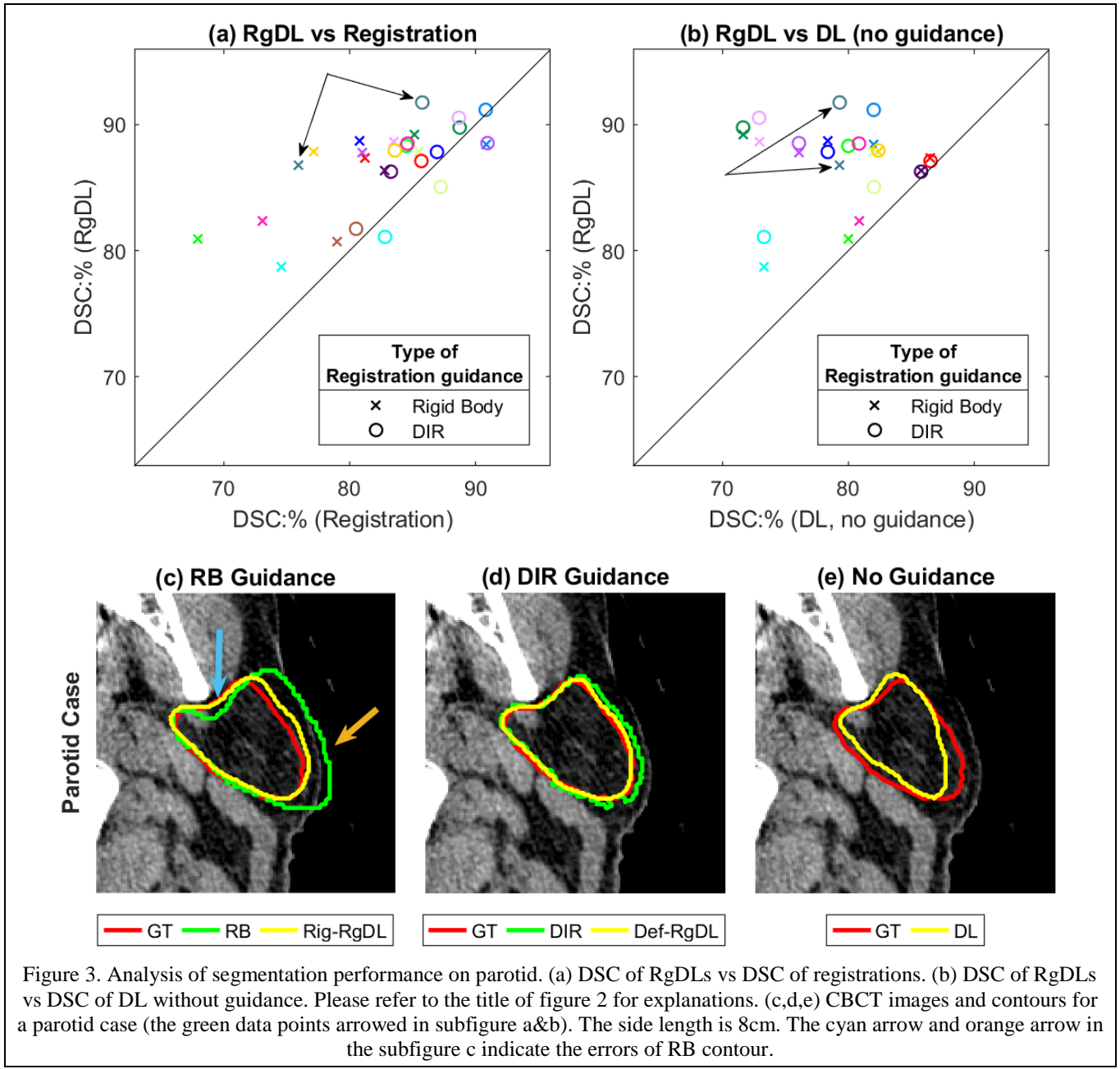

Figure 3. Analysis of segmentation performance on parotid. (a) DSC of RgDLs vs DSC of registrations. (b) DSC of RgDLs vs DSC of DL without guidance. Please refer to the title of figure 2 for explanations. (c,d,e) CBCT images and contours for a parotid case (the green data points arrowed in subfigure a&b). The side length is 8cm. The cyan arrow and orange arrow in the subfigure c indicate the errors of RB contour.

The DSC of RgDLs and baselines on 15 SMGs (left and right) in test set (table 1) are plotted in figure 4.a&b. The DSC of two RgDL implementations on 15 SMGs are consistently better than baseline methods as 24 out of 30 data points are above the diagonal line. The segmentation results of RgDLs and baselines on one SMG case (case A, arrowed in figure 4.a&b) are compared in figure 4.c-e. For this case, the accuracy of RgDLs in DSC (Rig-RgDL:67.8%, Def-RgDL:78.6%) are much higher than baselines (RB:51.5%,

DIR:73.9%, DL:62.4%), but the absolute values are still below 80%. On the images (figure 4.c-e), Rig-RgDL and Def-RgDL can refine the contours propagated by RB and DIR, though there are still some visible mismatch due to the vague boundary on CBCT. The segmentation results on another SMG case (case B, circled in figure 4.a&b) are compared in figure 4.f-h. For this case, the accuracy of RgDLs in DSC (Rig-RgDL:58.0%, Def-RgDL:53.8%) are comparable to registration (RB:54.3%, DIR:50.4%) but worse than DL (63.5%). Both RB (figure 4.f) and DIR (figure 4.g) falsely contoured a vessel as inside SMG, and RgDLs (yellow) failed to correct them. Direct DL segmentation managed to exclude the vessel, but still cannot be accurate on all boundaries due to the low image contrast.

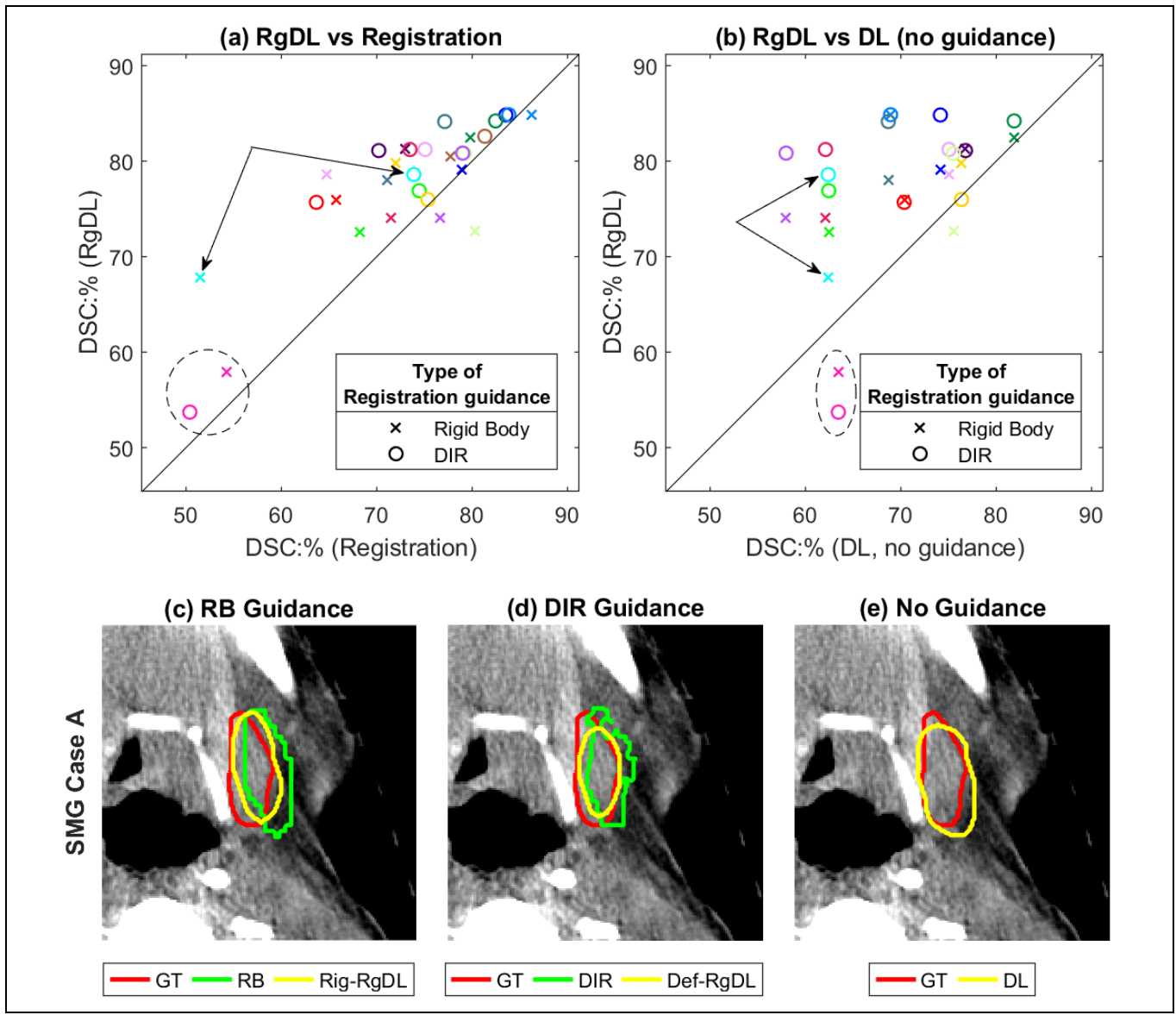

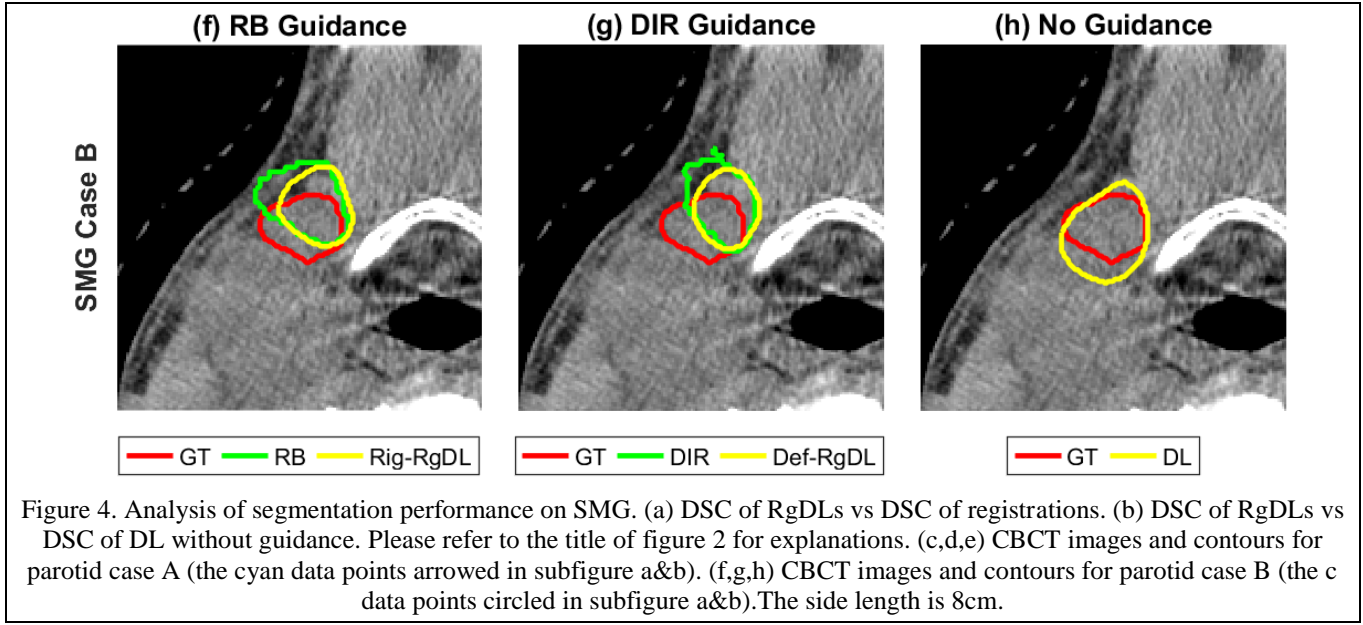

Figure 4. Analysis of segmentation performance on SMG. (a) DSC of RgDLs vs DSC of registrations. (b) DSC of RgDLs vs DSC of DL without guidance. Please refer to the title of figure 2 for explanations. (c,d,e) CBCT images and contours for parotid case A (the cyan data points arrowed in subfigure a&b). (f,g,h) CBCT images and contours for parotid case B (the c data points circled in subfigure a&b).The side length is 8cm.

The DSC of RgDLs and baselines on 14 masseters (left and right) in test set (table 1) are plotted in figure 5.a&b. The DSC of two RgDL implementations on 14 masseters are consistently better than baseline methods as 20 out of 28 data points are above the diagonal line. The segmentation results of RgDLs and baselines on one masseter case (arrowed in figure 5.a&b) are compared in figure 5.c-e. For this case, the accuracy of RgDLs in DSC (Rig-RgDL:90.2%, Def-RgDL:89.4%) are similar to registrations (RB:89.5%, DIR:88.9%) and much better than DL (78.3%) with large margins. As the masseter (muscular tissue) does not shrink or retract that much like parotid and SMG (glandular tissue), the guidance provided by RB or DIR are accurate in general position (green contours in figure 5.c&d), and the RgDL segmentations are accurate as well. The cyan arrows mark the artifact, which caused the over segmentation by DL (figure 5.e), when guidance was not provided.

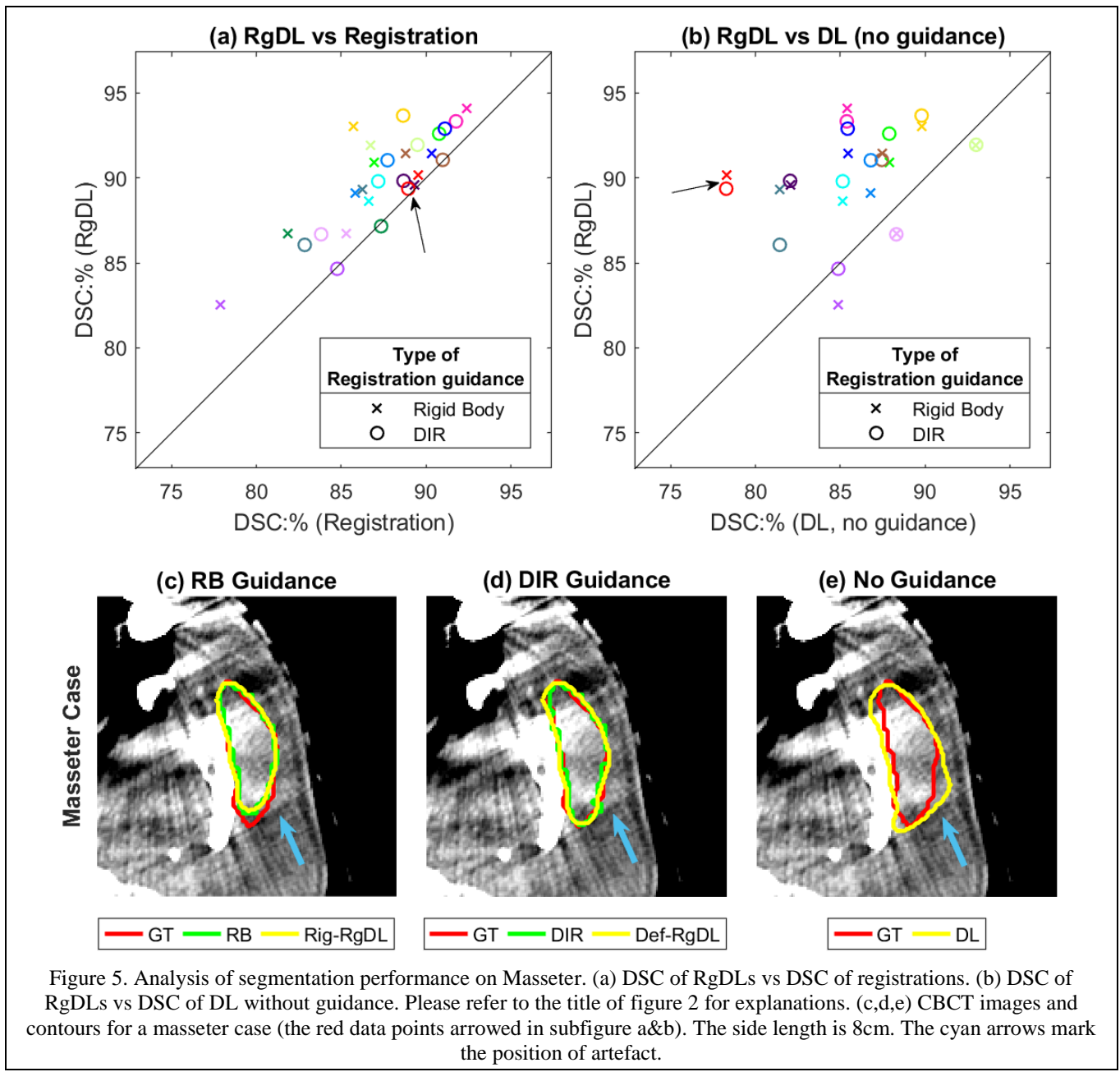

Figure 5. Analysis of segmentation performance on Masseter. (a) DSC of RgDLs vs DSC of registrations. (b) DSC of RgDLs vs DSC of DL without guidance. Please refer to the title of figure 2 for explanations. (c,d,e) CBCT images and contours for a masseter case (the red data points arrowed in subfigure a&b). The side length is 8cm. The cyan arrows mark the position of artefact.

The training time of each epoch is 110 seconds and total training time is about three hours (100 epochs). The inference time (forward propagation) of a single case is less than 0.05 seconds when DL model parameter has been loaded in GPU and computation graph has been constructed in previous runs. If we integrate RgDL into clinical ART workflow by using RgDL model as an online contour adaptation tool, the cost of time in addition to image registration and contour propagation will be RgDL model loading time (5.5 seconds) and inference time (1.3 seconds for the first run or 0.05 seconds after the first run).

### 3.2    Influence of the size of training dataset

We conducted further experiments to explore how the size of the training dataset influences RgDL. Without loss of generality, we chose Def-RgDL as the experiment subject as both implementations of RgDL function by utilizing spatial guidance and obtain positive results. We trained the Def-RgDL and DL (without guidance) model with four difference scenarios, increasing the size of training dataset from 5 to 10, 15, and 22. The trained Def-RgDL and DL models are still evaluated by the validation and test dataset in table 1. Their tendencies are compared to observe if using registration guidance can alleviate the training dataset size issue of DL.

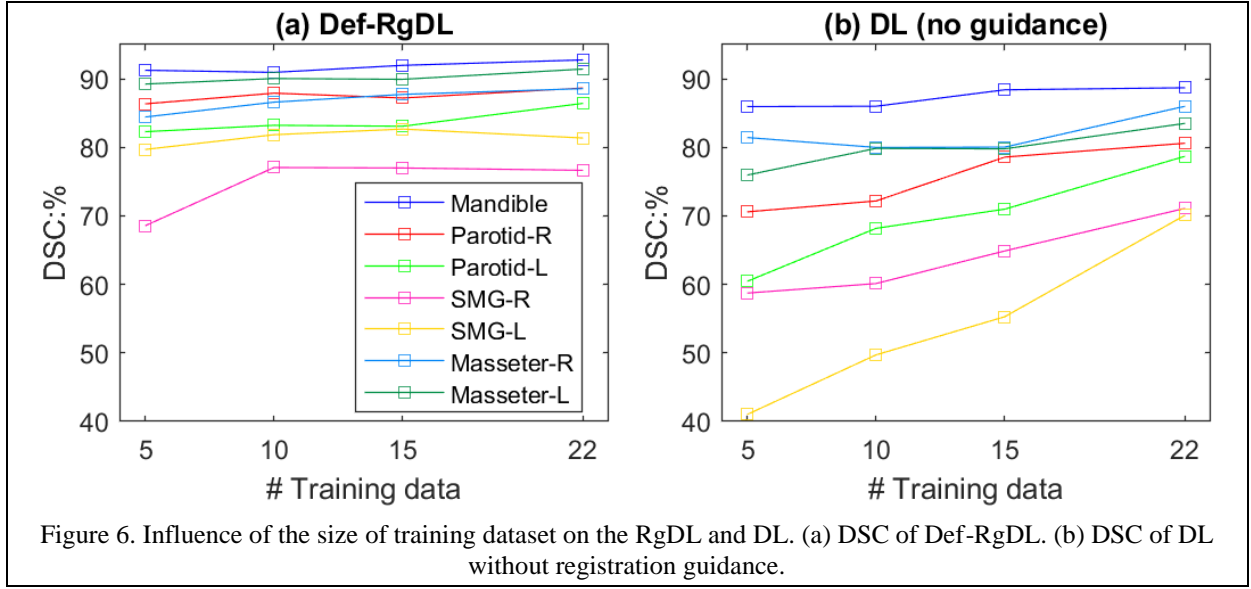

Figure 6. Influence of the size of training dataset on the RgDL and DL. (a) DSC of Def-RgDL. (b) DSC of DL without registration guidance.

The comparisons of mean DSCs with different numbers of training data are illustrated in figure 6. Overall, RgDL framework (Def-RgDL implementation, figure 6.a) outperformed DL without guidance (figure 6.b) by a large margin in all scenarios. Def-RgDL achieved DSCs of over 91.3%, 86.3%, 82.3%, 68.6%, 79.7%, 84.4% and 89.2% for Mandible, Parotid-R, Parotid-L, SMG-R, SMG-L, Masseter-R and Masseter-L, respectively, even with the smallest number (five) of training data. The DL model without guidance obtained inferior DSCs of 85.9%, 70.6%, 60.4%, 58.7%, 41.0%, 81.4% and 76.0% for Mandible, Parotid-R, Parotid-L, SMG-R, SMG-L, Masseter-R and Masseter-L respectively, when using only five patient cases for training.

The DSCs of DL (figure 6.b) improved substantially as the number of training data increased, especially for parotids and SMGs. The DSCs improved by 2.8%, 10.0%, 18.2%, 12.4%, 29.1%, 4.6% and 7.5% for Mandible, Parotid-R, Parotid-L, SMG-R, SMG-L, Masseter-R and Masseter-L, respectively, when the number of training data increased from 5 to 22. In contrast, RgDL (Def-RgDL, figure 6.a) showed stable segmentation performance with respect to changing the numbers of training dataset. When the number of training data decreased from 22 to 5, the DSC only decreased by 3.4% on average.

## 4    Discussion

In this work, we developed a RgDL framework to solve the challenge of segmenting HN organs on online CBCTs. Experiments on our in-house dataset demonstrated the promising segmentation capability of RgDL as compared with the baselines: registration contour propagations and direct DL segmentation. Our RgDL framework provides effective and efficient automatic OAR contouring on CBCT images for adaptive radiotherapy and requires only limited labeled data for model training.

To our knowledge, this RgDL framework is the first one that synergistically integrates image registration and DL segmentation model for HN OAR segmentation in online ART. Segmenting the online CBCT images can be approached from two directions: registration-based contour propagation or DL-based segmentation. Image registration takes advantage of the patient-specific prior—the positional information of organs in patient's planning images, but it has no general knowledge on organ segmentation. In contrast, DL-based segmentation methods can learn the patient anatomy and organ boundaries from population data, but the patient-specific prior information cannot be utilized by a trained model. Han[23] addressed the online CBCT segmentation (on abdomen OARs) by introducing population-based anatomy knowledge to image registration via a DL implementation. Our RgDL framework solves this task (on HN OARs) in a complementary perspective—incorporating patient-specific anatomy information to DL segmentation by registration guidance.

Our method can be regarded as refining registration-propagated contours via a DL method. Compared to the registration-based contour propagation, the DSC of Rig-RgDL and Def-RgDL are better than RB and DIR by 4.5% and 2.7%, respectively. From another perspective, the proposed method can be seen as an attention-guided DL method. With the assistance of registration-propagated contours, RgDL model can localize the latent boundary and extracts valuable representations that can promote segmentation of the actual contours on CBCT images. Because of the low soft-tissue contrast and the abundance of artifact and noise, DL model cannot precisely detect the boundaries of glands without guidance, which leads to its poor segmentation performance, especially on four glandular structures (parotids and SMGs). The DSC of DL on them is only 75.6% on average, 8.1% lower than Def-RgDL. By adding input channels for transformed / deformed masks in the first convolutional layer, the RgDL model adds only 6912 parameters—an increase of less than 0.1% over the single-channel-input DL model without guidance (~22M parameters)—to achieve this improvement in segmentation.

Canonical supervised learning in DL often requires a large amount of labeled data to train a robust neural network, but obtaining expert-labeled dataset is very expensive in medical fields. As shown in figure 6.b, the accuracy of DL segmentation model (without guidance) showed strong correlation to the size of training dataset. Because of the limited number of HN patients with available manual delineations on daily CBCTs, the optimal size of training data for the DL model to achieve its best performance have yet to be determined experimentally. From the curves shown in figure 6.b, it can be inferred that the basic U-Net would probably improve its segmentation performance if more CBCT labeled data were available for model training. This observation indicates the need for large labeled datasets to train a robust DL model. In comparison, the proposed RgDL framework can achieve accurate automatic segmentation of CBCT images with only limited training data. In fact, the segmentation results of the proposed method do not fluctuate much as the size of the training set changes. Therefore, this RgDL framework can be easily implemented with only dozens of labeled planning CT and CBCT images to achieve rapid and accurate HN organ segmentation for online ART.

We evaluated the segmentation accuracy of RgDL by using test cases in our in-house dataset, which are from the same source as the training and validation data. All CBCT images were acquired on Varian TrueBeam® onboard imaging system using the same machine setting (100 kVp / 150 mAs) and all CBCT contours were delineated by the same physician following our institutional protocol. As the quality of CBCT images and the contouring protocol may vary across different institutions, we recommend that each institution should commission their own model using their institutional data. RgDL only requires limited number of contoured CBCT, so data scarcity will become less of an issue.

Though our current study focused only on seven organs in HN, RgDL framework is applicable to organs in other anatomical sites, and even to the target volume. We believe that the RgDL framework will be more applicable to the patient-specific contours, such as CTV, PTV and tuning structures for planning. As contouring of those structures relies on patient-specific clinical information rather than image intensity itself so that the direct (population-based) deep learning has little clue to derive the right contours. However, such clinical prior knowledge has been encoded in the planning image contours and will be easily incorporated into online contouring by RgDL. As a promising research direction, we will investigate the RgDL framework for delineation of CTV and planning structure for online ART.

Moreover, RgDL framework has the flexibility to integrate other DL segmentation models and other registration algorithms ranged from traditional optimization-based algorithms to novel DL-based algorithms[23, 24]. Also, in this study, we only tested automatic segmentation on CBCT images. Many other image modalities, such as MRI, can be fitted into this framework, particularly for the application in MR-Linac based online ART[25]. In the future, we will work to extend the implementations of RgDL to the aspects mentioned above.

## 5    Conclusion

In this study, we developed a novel registration-guided DL (RgDL) segmentation framework that effectively mitigates the obstacles caused by the poor quality of CBCT images and the demand for numerous labeled training data in DL. Experimental results demonstrated promising segmentation performance for seven HN OARs with RgDL, which can potentially be applied in online ART.

## 6    Acknowledgements

We would like to thank Jonathan Feinberg for editing the manuscript.